\documentclass[conference,a4paper]{APSIPA2025}
\usepackage{amsmath}
\usepackage{graphicx}
\usepackage{multirow}
\usepackage{threeparttable}
\usepackage{booktabs}
\usepackage{amssymb}
\usepackage{tabularx}
\usepackage{caption}

\usepackage{geometry}
\usepackage{fancyhdr}
\usepackage{url}

\fancypagestyle{firststyle}{
  \fancyhf{}
  \fancyhead[C]{2025 Asia Pacific Signal and Information Processing Association Annual Summit and Conference (APSIPA ASC)}
}

\begin{document}

\title{Pre-training Autoencoder for Acoustic Event Classification via Blinky}

\author{
  \authorblockN{
    Xiaoyang Liu and Yuma Kinoshita
  }
  \authorblockA{
    Tokai University, Japan
  }
}

\maketitle
\thispagestyle{firststyle}
\pagestyle{fancy}
\cfoot{ }

\begin{abstract}
    In the acoustic event classification (AEC) framework that employs Blinkies, audio signals are converted into LED light emissions and subsequently captured by a single video camera. However, the 30 fps optical transmission channel conveys only about 0.2\% of the normal audio bandwidth and is highly susceptible to noise.
    We propose a novel sound-to-light conversion method that leverages the encoder of a pre-trained autoencoder (AE) to distill compact, discriminative features from the recorded audio.
    To pre-train the AE, we adopt a noise-robust learning strategy in which artificial noise is injected into the encoder’s latent representations during training, thereby enhancing the model’s robustness against channel noise.
    The encoder architecture is specifically designed for the memory footprint of contemporary edge devices such as the Raspberry Pi 4.
    In a simulation experiment on the ESC-50 dataset under a stringent 15 Hz bandwidth constraint, the proposed method achieved higher macro-F$_1$ scores than conventional sound-to-light conversion approaches.
\end{abstract}

\section{Introduction}

Acoustic event classification (AEC) is the task of estimating the posterior probabilities of multiple predefined event categories from an observed acoustic signal and outputs the most probable category.
This technology is widely adopted in applications such as environmental monitoring, intelligent surveillance, and home automation.
Significant progress in single-channel AEC has been achieved through the development of deep neural network (DNN)-based methods.
In such methods, acoustic features like time-frequency representations are first extracted from raw audio signals and then provided as input to DNN models (e.g., CNNs and Transformers) to obtain class-posterior probabilities~\cite{kong2020pannslargescalepretrainedaudio,chen2022htsathierarchicaltokensemanticaudio, gong2021astaudiospectrogramtransformer}.
In addition, considerable efforts have been devoted to implementing AEC methods on edge devices \cite{huang2024tinychirpbirdsongrecognition,Mohaimenuzzaman_2023_acdnet}.

Besides spectral information from a single microphone, spatial information from a distributed array~\cite{zhang2024soundeventlocalizationclassification,stowell_detection_2015} improves AEC.
Placing several microphones near sound sources raises the signal-to-noise ratio, and the distinct inter-microphone delays preserve clues needed to separate simultaneously occurring events.
Fully exploiting these spatial information demands sample-level synchronization across channels. Because wireless microphones are inherently asynchronous, recent work has focused on blind time alignment~\cite{cherkassky_blind_2017}.
Wireless transmission also restrict throughput: at 16 kHz/16-bit resolution each sensor yields approximately 32 kB of data per second, so many microphones can conflict with the available communication bandwidth and storage resources.

Addressing these issues, a sound-to-light conversion device named a Blinky has been developed~\cite{scheibler_blinkies_2020, scheibler_blinkies_2018,ishii_real-time_2021, nishida_estimation_2022}.
These Blinkies can convert acoustic signals into varying intensities
of light via an inbuilt light-emitting diode (LED).
A video camera is employed to synchronously
record LED brightness from multiple Blinkies spread
over a wide region.
Aggregating the Blinky signals from the recorded video,
the fusion center, which is a high-performance server, performs AEC by integrating and analyzing acoustic information.
Previous studies have validated its effectiveness in sound localization, speech enhancement, \cite{scheibler_blinkies_2018} and AEC tasks \cite{kinoshita_end--end_2021, kinoshita_e2e_2024, kinoshita2021analysis}, demonstrating the feasibility of optical transmission of acoustic features.

However, Blinky remains constrained by severe bandwidth limitations imposed by the camera's frame rate, which is typically 30 frames per second (FPS), as well as by noise introduced during light-signal transmission through the air.
This frame rate is considerably lower than the standard sampling rate of microphones, which is typically 16 kHz, and the bandwidth available for LED signals is approximately 1/533 that of audio signals.
In an initial attempt to transmit sound signals via LED light signals,
sound power was simply used as the basis for sound-to-light conversion; however, this approach is suboptimal for AEC due to restricted bandwidth and noise susceptibility.
For this reason, Kinoshita et al. have proposed training
a DNN for sound-to-light conversion in an end-to-end manner.
Nevertheless, Kinoshita's method requires backpropagation of the loss of AEC through the unknown, non-differentiable physical light transmission channel, rendering it impractical for training the DNN on real-world environments.
Therefore, a critical challenge arises: how to transmit semantically discriminative features through severely bandwidth-limited channels while preserving Blinky’s practical deployment advantages.

To overcome this challenge, in this paper, we propose a novel sound-to-light conversion method that does not require backpropagation through the physical transmission channel.
For the proposed method,
we first pre-train a lightweight autoencoder (AE) to encode audio signals into compact latent vectors while preserving their essential information, where artificial noise is injected into the latent vectors during training to enhance robustness against noise.
The encoder network of the pre-trained AE is then deployed on each Blinky, enabling the extraction of latent vectors from the recorded audio signals.
The latent vectors are subsequently transmitted as light signals using Blinky's four LEDs.
AEC is performed at a centralized fusion center, where a DNN classifier analyzes the LED signals captured by a camera.

We evaluated the efficacy of our AE-based sound-to-light conversion for acoustic event classification (AEC) through simulation studies utilizing the ESC-50 dataset.
The experimental results demonstrate that our method outperforms both sound-power-based and end-to-end training approaches in terms of macro-F1 score, improving performance from 0.34 (sound-power-based) and 0.31 (end-to-end training) to 0.54.

Our primary contributions are as follows:
\begin{itemize}
    \item We introduce a novel pre-trained AE-based sound-to-light conversion method, along with a noise-robust training strategy for AE pre-training, for acoustic sensing with Blinkies, enabling the transmission of discriminative features over noisy channels under the stringent bandwidth constraint of 15 Hz.
    \item We present a novel AE architecture whose encoder has a total inference-time memory footprint of approximately 3.40 MB, which is well within the capabilities of contemporary edge platforms such as the Raspberry Pi 4.
    \item We provide an open-source implementation of our simulation experiments, including both the proposed and conventional sound-to-light conversion methods.
\end{itemize}
\section{Problem Setting}
\label{sec:problem_setting}

We address the challenge of implementing the AEC framework using Blinkies. This section formalizes the AEC pipeline and defines the core research problem.

\subsection{Formalization of AEC Framework using Blinkies}
The AEC procedure using $N$ Blinkies consists of three sequential stages: on-device embedding ($\mathrm{Embed}(\cdot)$), optical signal transmission ($\mathrm{Transmit}(\cdot)$), and downstream classification ($\mathrm{Classify}(\cdot)$). This pipeline is expressed as:
\begin{align}
    \boldsymbol{z}_i  & = \mathrm{Embed}(\boldsymbol{x}_i),\label{eq:ps_embedding}                               \\
    \boldsymbol{z'}_i & = \mathrm{Transmit}(\boldsymbol{z}_i), \label{eq:ps_channel}                             \\
    \hat{y}           & = \mathrm{Classify}(\boldsymbol{z'}_1, \cdots, \boldsymbol{z}_N). \label{eq:ps_classify}
\end{align}
\begin{figure*}[ht]
    \centering
    \includegraphics[width=\textwidth]{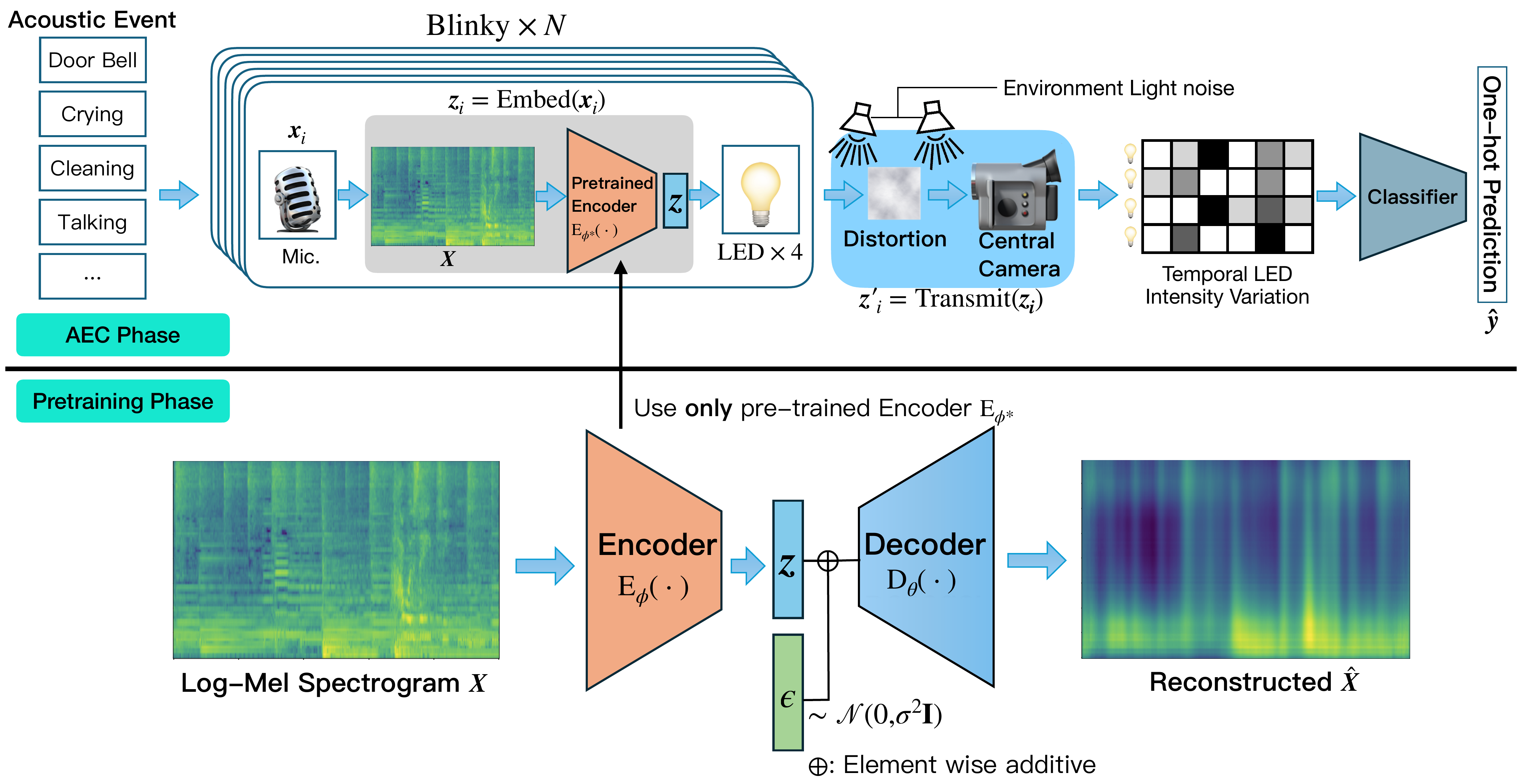}
    \caption{Overview of the proposed framework.
        \textbf{Pre-training phase} (bottom): An autoencoder (AE) is pre-trained to learn noise-robust representations; Gaussian noise ($\boldsymbol{\epsilon}$) is injected into the latent vector ($\boldsymbol{z}$) and the decoder ($D_{\theta}$) reconstructs the original log-Mel spectrogram ($\boldsymbol{X}$).
        \textbf{AEC phase} (top): At inference, the frozen encoder ($E_{\phi}^{*}$) produces a compact latent representation that modulates four LEDs. A central camera records the signal as \emph{temporal LED intensity variations}, and the downstream classifier predicts the event label ($\hat{\boldsymbol{y}}$).}
    \label{fig:proposed_system_diagram}
\end{figure*}
As illustrated in Fig.~\ref{fig:proposed_system_diagram},
the $i$-th Blinky first acquires an acoustic signal $\boldsymbol{x}_i$.
The embedding function $\mathrm{Embed}(\cdot)$ on the Blinky maps the captured signal to a feature vector $\boldsymbol{z}_i \in \mathbb{R}^{L}$, which subsequently modulates the device's LEDs.
The resulting light signal propagates through the environment and is captured by a central camera;
the composite process is abstracted by the function $\mathrm{Transmit}(\cdot)$.
Finally, the classifier $\mathrm{Classify}(\cdot)$ operates on the collection of received signals $\{\boldsymbol{z'}_i\}$ to infer the acoustic event label $\hat{y}$.

Within the transmission channel, Blinkies' LED signals are perturbed by ambient illumination and noise before being temporally resampled by the camera.
Accordingly, we model the channel as the composition of a distortion operator followed by camera resampling, i.e., $\mathrm{Transmit}(\cdot) = (\mathrm{Resample} \circ \mathrm{Distort}(\cdot))$.
The distortion is parameterized as follows:
\begin{align}
    \mathrm{Distort}(\boldsymbol{z}) & = a\boldsymbol{z} + b\boldsymbol{1} + \boldsymbol{\epsilon}, \quad \boldsymbol{\epsilon} \sim \mathcal{N}(0, \sigma^2\mathbf{I}), \label{eq:ps_distort}
\end{align}
where $a$ denotes the attenuation coefficient---inversely proportional to the squared distance between the camera and a Blinky---and $b$ captures the bias introduced by ambient light,
whereas $\boldsymbol{\epsilon}$ represents additive white Gaussian noise with covariance $\sigma^2 \boldsymbol{I}$.
The vector $\boldsymbol{1}$ is the $L$-dimensional all-ones vector
and the matrix $\boldsymbol{I}$ is the identity matrix with a size of $L \times L$.
The resampling operator $\mathrm{Resample}: \mathbb{R}^{L} \rightarrow \mathbb{R}^{N_{\text{LED}} \times \lfloor F_s \times T \rfloor}$ encapsulates the camera's temporal sampling, governed by its frame rate $F_s$, the number of LEDs per device $N_{\text{LED}}$, and the audio segment duration $T$.
To respect the sensor's dynamic range,
resampled values are clipped to the interval $[0, 1]$.

\subsection{Challenge: Designing Embedding Function}
Within this framework, the central challenge is to design an embedding function $\mathrm{Embed}(\cdot)$ that retains as much acoustic information as possible despite stringent bandwidth constraints and pronounced channel noise.
To date, two principal strategies have been explored.

\paragraph{Sound Power}
The simplest method defines $\mathrm{Embed}(\cdot)$ by mapping the raw waveform $\boldsymbol{x}(t)$ to its power: $\boldsymbol{z} = (x(1)^2, x(2)^2, \cdots, x(T)^2)^\top$.
Although computationally negligible, the downsampling process in the camera discards most of the semantic information (e.g., timbre),
resulting in insufficient performance on AEC tasks.

\paragraph{End-to-End Training Approach}
To convey more discriminative features, Kinoshita et al.~\cite{kinoshita_end--end_2021} proposed using a DNN encoder for $\mathrm{Embed}(\cdot)$ and optimized both $\mathrm{Embed}(\cdot)$ and $\mathrm{Classify}(\cdot)$ (Eqs.~\ref{eq:ps_embedding} and \ref{eq:ps_classify}) in an end-to-end fashion.
However, this strategy is impractical in real-world deployments:
the transmit channel (Eq.~\ref{eq:ps_channel}) constitutes an unknown, non-differentiable physical process, thereby blocking gradient flow back to the on-device encoder.

Both approaches therefore exhibit critical shortcomings: the sound-power approach is too rudimentary for high-accuracy classification, whereas the end-to-end training approach is physically unrealizable. Consequently, a practical solution is required---one that permits the transmission of expressive, learned features without necessitating differentiation through the physical channel.

\section{Methodology}
\label{sec:methodology}

To overcome the limitations described in the previous section, we propose a practical solution that decouples the feature embedding from the non-differentiable physical channel.
Our approach, illustrated in Fig.~\ref{fig:proposed_system_diagram}, realizes the embedding function $\mathrm{Embed}(\cdot)$ via the encoder of a pre-trained autoencoder (AE), denoted as $E_{{\phi}^{*}}(\cdot)$.

The procedure begins by transforming the raw input waveform $\boldsymbol{x}$ into a log-Mel spectrogram $\boldsymbol{X}=\mathrm{Logmel}(\boldsymbol{x})$,
a widely adopted time-frequency representation of audio.
The embedding function then maps the waveform to a latent vector $\boldsymbol{z}$ through the log-Mel representation:
\begin{equation}
    \boldsymbol{z} = \mathrm{Embed}(\boldsymbol{x}) = E_{{\phi}^{*}}(\boldsymbol{X}).
    \label{eq:embed_ae}
\end{equation}
By pre-training the encoder,
we obtain a compact yet discriminative representation that is suited for transmission,
thereby obviating the need for end-to-end optimization through the physical channel.
This section details the AE's learning objective, its network architecture, pre-training procedure, and training procedure for the downstream classifier.
\footnote{All training scripts, including Autoencoder pre-training and downstream evaluation, are available at \url{https://github.com/ykinolab-tokai/multi-ae-ace}.}

\subsection{Learning Objective}
The primary objective of an AE is to distill a compact representation from a large, unlabeled audio dataset (e.g. Google AudioSet \cite{gemmeke_audio_2017}) by means of a self-supervised reconstruction task.
A standard AE minimizes the reconstruction error between an input log-Mel spectrogram $\boldsymbol{X}$ and its reconstruction $\hat{\boldsymbol{X}}$ via
\begin{equation}
    \mathcal{L}_{\text{rec}} = \mathbb{E}\left[ ||\boldsymbol{X} - D_\theta(E_\phi(\boldsymbol{X}))||^2_F \right],
    \label{eq:recon_loss}
\end{equation}
where $E_\phi(\cdot)$ and $D_\theta(\cdot)$ represent the encoder and decoder, respectively;
the optimized parameters after pre-training are $(\phi^*, \theta^*)$.

However, this objective does not consider the perturbation of the latent vector $E_\phi(\boldsymbol{X})$ by channel noise.
To address this issue, we adopt a noise-robust training strategy:
Gaussian noise $\boldsymbol{\epsilon}$ is injected into the latent vectors during training to mimic channel distortion and compel the model to produce a more resilient representation.
The revised objective is
\begin{equation}
    \mathcal{L}_{\text{rec-robust}} = \mathbb{E}\left[ ||\boldsymbol{X} - D_\theta(E_\phi(\boldsymbol{X}) + \boldsymbol{\epsilon})||^2_F \right], \quad \boldsymbol{\epsilon} \sim \mathcal{N}(0, \sigma^2\mathbf{I}),
    \label{eq:recon_loss_robust}
\end{equation}
thereby encouraging $E_\phi(\cdot)$ to produce channel-invariant and discriminative features.
\subsection{AE Architecture}
\label{subsec:ae_architecture_rationale}
\begin{table}[t]
    \centering
    \caption{
        Autoencoder architecture.
        Output shapes are listed as $\mathrm{frequency} \ F \times \mathrm{time} \ T$.
        Self-Attention block consists of multi-head attention layer with two heads, layer normalization, feed-forward network (FFN) that expands feature dimension from 40 to 128 and then projects it back, and layer normalization with residuals.
        $k$, $s$, and $c$ denotes kernel size, stride, and channel,
        respectively.
    }
    \label{tab:ae_architecture}
    \begin{tabular}{l c}
        \toprule
        \textbf{Layer}                                & \textbf{Output Shape} ($F \times T$) \\
        \midrule
        \textit{Input: Log-Mel Spectrogram}           & [80, 501]                            \\
        \midrule
        \multicolumn{2}{l}{\textbf{Encoder} ($E_\phi$)}                                      \\
        Conv1d (c=40, k=5, s=1), ReLU                 & [40, 497]                            \\
        Self-Attention (embed=40, heads=2)$^*$        & [40, 497]                            \\
        ReLU                                          & [40, 497]                            \\
        Conv1d (c=20, k=7, s=2), ReLU                 & [20, 246]                            \\
        Conv1d (c=20, k=11, s=3), ReLU                & [20, 79]                             \\
        Flatten                                       & [1, 1580]                            \\
        Linear (in\_features=1580, out\_features=300) & [1, 300]                             \\
        \midrule
        \multicolumn{2}{l}{\textbf{Decoder} ($D_\theta$)}                                    \\
        Linear (in\_features=300, out\_features=1580) & [1, 1580]                            \\
        Reshape                                       & [20, 79]                             \\
        ConvTranspose1d (c=20, k=11, s=3), ReLU       & [20, 245]                            \\
        ConvTranspose1d (c=40, k=7, s=2), ReLU        & [40, 495]                            \\
        ConvTranspose1d (c=80, k=5, s=1)              & [80, 499]                            \\
        Interpolate (target\_time\_dim=501)           & [80, 501]                            \\
        \midrule
        \textit{Output: Reconstructed Log-Mel}        & [80, 501]                            \\
        \bottomrule
    \end{tabular}
\end{table}

The Encoder architecture, summarized in Table~\ref{tab:ae_architecture}, is a hybrid design that combines convolutional neural networks (CNNs) and self-attention.
A 1-D convolutional layer first extracts local temporal patterns, after which a multi-head self-attention layer captures long-range dependencies, achieving a global receptive field without excessive network depth.
A series of strided convolutional layers performs temporal downsampling before projecting the signal into an $L$-dimensional latent space.
We choose $L$ to match the physical channel capacity, i.e.,
$L = 15 \, \mathrm{symbol/s} \times 4 \, \mathrm{LEDs} \times 5 \mathrm{s} = 300$, where $15$\,Hz is the per-LED Nyquist rate imposed by the 30\,FPS camera.
Unlike audio transformer models~\cite{gong2021astaudiospectrogramtransformer}, which prioritize accuracy without reducing bitrate, our AE explicitly compresses the representation to meet the optical-channel capacity.

A crucial deployment constraint is the encoder's computational footprint on resource-constrained hardware.
Our encoder contains 517,508 trainable parameters,
occupying roughly 2.04\,MB of memory,
and its total inference-time footprint is about 3.40\,MB.
This is comfortably within the capacity of modern edge platforms such as the Raspberry Pi 4.

\subsection{Pre-training}
\label{subsec:ae_preprocessing}
We aim to infer a single acoustic-event label $y$ from a set of 5-second audio waveform $\{ \boldsymbol{x}_i \}$ sampled at 16\,kHz.
Each 10-second audio clip from the Google AudioSet~\cite{gemmeke_audio_2017} is first down-mixed to monaural and resampled to 16\,kHz.
For data augmentation, a 5-second segment is then randomly cropped from every clip.
A log-Mel spectrogram $\boldsymbol{X}$ is then computed from this segment, using a Hann window with a length of 400, a 160-sample hop length, and 80 Mel filters.

The AE is pre-trained for 300 epochs on the balanced subset of AudioSet with the Adam optimizer~\cite{kingma_adam_2017}, an initial learning rate of $1 \times 10^{-3}$ and a batch size of 1280.
A \texttt{ReduceLROnPlateau} learning rate scheduler lowers the learning rate by a factor of 0.1 whenever the validation loss fails to improve for 10 consecutive epochs.

\subsection{Training Downstream Classifier}
\label{subsec:classifier_training}
To train the downstream classifier that predicts acoustic event labels
from the transmitted data, we first record the Blinky signals corresponding to each input audio instance in the training set
with the central camera, thereby obtaining a set of latent vectors $\{\boldsymbol{z}'_i\}$ (see Eqs.~\ref{eq:ps_embedding} and \ref{eq:ps_channel}).
The classifier is subsequently trained to infer the target labels $y$
from these latent representations,
whereas the parameters of the pre-trained encoder $E_{{\phi}^{*}}$ are kept fixed throughout this phase.
\section{Experiments}
\label{sec:experiments}

We evaluated the efficacy of our AE–based sound-to-light conversion for AEC through a simulation study.

\subsection{Experimental Setup}
\label{subsec:exp_setup}

The experiments were conducted in a simulated rectangular room measuring $8 \times 6 \times 4\,\text{m}^{3}$, where five Blinkies were deployed.
The process of estimating the class label $\hat{y}$ from the waveform $\boldsymbol{x}_i$ received by the $i$-th Blinky follows the procedure
outlined in Eqs.~\ref{eq:ps_embedding} to \ref{eq:ps_classify}.
AEC performance was measured using the $F_{1}$ score
and model selection relied on the highest $F_{1}$ score obtained on a validation set.

To assess the contribution of our AE-based sound-to-light conversion,
we compared the following configurations for $\mathrm{Embed}(\boldsymbol{x}_i)$:
\begin{itemize}
      \item \textbf{Log-Mel spectrogram without embedding}:
            Converts the raw waveform $\boldsymbol{x}$ into a log-Mel spectrogram $\boldsymbol{X}$, with no further embedding, i.e., $\mathrm{Embed}(\boldsymbol{x})=\boldsymbol{X}$. This evaluation is provided as a reference for the ideal classification performance on the same data, without the Blinky channel constraints.
      \item \textbf{Sound Power}:
            Computes $\boldsymbol{z}$ as the signal power, i.e., $\mathrm{Embed}(\boldsymbol{x})=(x(1)^2, \cdots, x(T)^2)^\top$, following the approach in \cite{scheibler_blinkies_2020}.
      \item \textbf{End-to-End}:
            Computes $\boldsymbol{z}$ by using a DNN encoder $E_{\phi}(\cdot)$, i.e., $\mathrm{Embed}(\boldsymbol{x}) = E_{\phi}(\boldsymbol{X})$. In this method, both the encoder and classifier are jointly optimized in an end-to-end fashion
            using the classification loss.
            The encoder architecture is identical to that of our pre-trained encoder $E_{\phi}$.
      \item \textbf{Autoencoder}:
            Computes $\boldsymbol{z}$ as the latent representation produced by an autoencoder pre-trained with Eq. \ref{eq:recon_loss}, i.e.,
            $\mathrm{Embed}(\boldsymbol{x})=E_{{\phi}^{*}}(\mathrm{Logmel}(\boldsymbol{x}))$.
      \item \textbf{Noise-robust autoencoder (Ours)}:
            Computes $\boldsymbol{z}$ as the latent representation from our noise-aware autoencoder pre-trained with Eq.~\ref{eq:recon_loss_robust},
            i.e., $\mathrm{Embed}(\boldsymbol{x})=E_{{\phi}^{*}}(\mathrm{Logmel}(\boldsymbol{x}))$.
\end{itemize}

We utilized the ESC-50 dataset~\cite{esc50_2015}, partitioned into training, validation, and test sets with an 8:1:1 ratio
on a per-class basis to ensure class-wise balance across all subsets.
The acoustic waveform $\boldsymbol{x}_i$ received at the $i$-th Blinky was generated by convolving a source signal $s$ with the corresponding room impulse response (RIR) between that source and the Blinky.
To simulate class-dependent spatial characteristics,
a single sound source was randomly positioned for each class, with the $k$-th source assigned to the $k$-th class.
Source and Blinky positions remained fixed throughout the experiment.
RIRs were generated using the image-source method implemented in the \textit{pyroomacoustics}~\cite{Scheibler_pyroom_2018},
with a wall-absorption coefficient of 0.4 and a maximum image source order of 10, then zero-padded to equal length.
Parameter $a$, $b$, and $\sigma$ of Eq.~\ref{eq:ps_distort} were $1$, $0.1$, and $0.05$, respectively.

The classifier $\mathrm{Classify}(\cdot)$ was implemented as a ResNet-18~\cite{he_deep_2015}.
During training the classifier,
we applied BC-learning~\cite{tokozume_bclearning_2018}
as a data augmentation,
wherein two samples from different classes and their one-hot labels were linearly combined to produce mixed inputs and soft targets.
The classifier was trained using the Kullback–Leibler (KL) divergence
as the loss function and Adam optimizer~\cite{kingma_adam_2017}.
As in the pre-training phase, we used the \texttt{ReduceLROnPlateau} scheduler in the classifier training phase, initializing the learning rate of $1 \times 10^{-2}$. The scheduler monitored the validation loss and reduced the learning rate by a factor of 0.5 whenever the validation loss plateaued.

This simulation experiment was implemented using the \textit{PyTorch} framework~\cite{paszke_2019_pytorch}.

\subsection{Results}
\label{subsec:results}
\begin{table*}[h]
    \centering
    \caption{Macro-averaged F$_1$ score (mean $\pm$ std. across four independent trials) on test set under three transmission conditions:
        no degradation $\mathrm{Transmit}(\boldsymbol{z}) = \boldsymbol{z}$ (\textit{Ideal channel}),
        resampling $\mathrm{Transmit}(\boldsymbol{z}) = \mathrm{Resample}(\boldsymbol{z})$ (\textit{Resample}),
        and resampling with distortions
        $\mathrm{Transmit}(\boldsymbol{z}) = (\mathrm{Resample} \circ \mathrm{Distort})(\boldsymbol{z})$ (\textit{Resample + Distort}). Definitions of each embedding method are provided in Section~\ref{subsec:exp_setup}.}
    \label{tab:exp_results_f1}
    \begin{tabular}{l c c c}
        \toprule
        \textbf{Embedding method}                      & \textbf{Ideal channel}       & \textbf{Resample}            & \textbf{Resample + Distort}  \\
        \midrule
        \textit{Log-Mel spectrogram without embedding} & 0.9949                       & ---                          & ---                          \\
        \textit{Sound Power}                           & ---                          & 0.3976 $\pm$ 0.0189          & 0.3359 $\pm$ 0.0238          \\
        \textit{End-to-End}                            & ---                          & 0.4965 $\pm$ 0.0731          & 0.3077 $\pm$ 0.0705          \\
        \textit{Autoencoder}                           & 0.6842 $\pm$ 0.0448          & 0.6696 $\pm$ 0.0109          & 0.5227 $\pm$ 0.0206          \\
        \textit{Noise-Robust Autoencoder (Ours)}       & \textbf{0.7248 $\pm$ 0.0405} & \textbf{0.7279 $\pm$ 0.0201} & \textbf{0.5357 $\pm$ 0.0178} \\
        \midrule
        \textit{ESC-50 Human Benchmark}                & \multicolumn{3}{c}{0.8130}                                                                 \\
        \bottomrule
    \end{tabular}
\end{table*}

Table~\ref{tab:exp_results_f1} presents macro-averaged $F_{1}$ scores on the test set under three transmission scenarios: (i) no degradation, $\mathrm{Transmit}(\boldsymbol{z})=\boldsymbol{z}$; (ii) resampling only, $\mathrm{Transmit}(\boldsymbol{z})=\mathrm{Resample}(\boldsymbol{z})$; and (iii) resampling followed by optical distortions, $\mathrm{Transmit}(\boldsymbol{z})=(\mathrm{Resample}\circ\mathrm{Distort})(\boldsymbol{z})$.
Except for the log-Mel spectrogram without embedding, each method was evaluated four times with different random seeds, and the mean and standard deviation across these four runs are reported in Table~\ref{tab:exp_results_f1}.

Across all conditions, our AE-based method substantially outperforms both the sound-power and the end-to-end approaches.
Moreover, our noise-robust autoencoder achieves higher $F_{1}$ scores than the standard autoencoder,
highlighting the effectiveness of the noise-aware training objective described in Eq.~\ref{eq:recon_loss_robust}.

\subsection{Discussion and Limitations}
Through simulation experiments, we demonstrated the efficacy of the proposed noise-robust autoencoder for sound-to-light conversion. Nevertheless, because these experiments were purely virtual, its performance under real-world conditions remains unverified.
Our immediate objective is therefore to implement the framework in a physical environment and conduct a comprehensive evaluation of its AEC accuracy.
This assessment will also quantify the computational overhead and determine the feasibility of real-time processing on embedded hardware.

Moreover, the present study does not incorporate the spatial characteristics inherent in audio signals during AE training. Consequently, the latent vectors may lack room-impulse response (RIR) information contained in the original signal,
potentially diminishing the spatial cues available for AEC.
Designing a loss function that explicitly preserves spatial information could yield richer embeddings and further improve performance.

Another limitation of this study is that our evaluation focuses exclusively on single-event clips and does not address scenarios in which two or more acoustic events occur simultaneously. This work primarily emphasizes validating whether the proposed method functions as intended, rather than exhaustively delineating its operational boundaries; therefore, multi-event classification is deferred to future investigations. Nevertheless, since the downstream classifier is trained with BC learning—thereby being exposed to mixtures and soft labels—the resulting decision function is capable of handling mixed inputs to some extent, indicating a promising potential for the recognition of concurrent events in more complex environments.

It is further assumed that each Blinky transforms its recorded audio into a low-dimensional vector within a shared 5-second time window, and it is not assumed that recordings occur over differing temporal intervals across devices. However, this assumption may not hold in practical deployments, where device-specific asynchrony can arise. Consequently, the design of robust synchronization mechanisms and the development of training or inference strategies resilient to temporal misalignment represent important avenues for future research.
Finally, although we employed an autoencoder for unsupervised representation learning, alternative methods---such as generative approaches (e.g., variational autoencoders) and contrastive learning objectives---warrant investigation.
\section{Conclusion}
\label{sec:discussion_conclusion}
In this paper, we propose a practical sound-to-light conversion method for AEC using Blinkies.
Conventional methods either suffer from low accuracy or face difficulties in real-world deployment due to non-differentiable physical channels.
To overcome this, we separate the training of the feature embedding function from the physical channel and use a pre-trained autoencoder (AE) encoder for the embedding.
For pre-training AE, a noise-robust training strategy is introduced to enhance robustness against distortion in the transmission channel.
Simulation results show that the proposed AE-based method consistently outperforms conventional approaches across various conditions, achieving higher F$_1$ scores.

Despite promising results in simulations, real-world evaluation remains a future task.
Additionally, incorporating spatial information into the training objective and exploring other self-supervised methods like VAE or contrastive learning are potential future directions.
\section*{Acknowledgment}
This work was supported by JSPS KAKENHI Grant Number JP22K17915.

\end{document}